# A universal alphabet and rewrite system

Peter Rowlands[1] and Bernard Diaz[2]

[1] Science Communication Unit, Department of Physics
[2] Department of Computer Science
The University of Liverpool
Peach Street, Liverpool, UK, L69 7ZF

e-mail: p.rowlands@liv.ac.uk, b.m.diaz@liv.ac.uk

**Abstract**. *We present two ways in which an infinite universal alphabet may be generated using a novel rewrite system that conserves zero (a special character of the alphabet and the symbol for that character) at every step. The recursive method delivers the entire alphabet in one step when invoked with the zero character as the initial subset alphabet. The iterative method with the same start delivers characters that act as ciphers for properties that the developing subset alphabet contains. These properties emerge in an arbitrary sequence and there are an infinite number of ways they may be selected. The subset alphabets in addition to having mathematical interpretation as algebra can also be constrained to emerge in a minimal way which then has application as a foundational physical system. Each subset alphabet may itself be the basis of a rewrite system where rules that operate on symbols (representing characters) or collections of symbols manipulate the specific properties in a dynamic way.*

## 1   Introduction

Rewrite systems are synonymous with computing in the sense that most software is written in a language that must be rewritten as symbols for some hardware to interpret. Formal rewrite, or production, systems are pieces of software that take an object usually represented as a string of characters and using a set of *rewrite rules* generate a new string representing an altered state of the object. If required, a second *realisation system* takes the string and produces a visualisation or manifestation of the objects being represented.

We seek to extend the applicability and power of rewriting by examining how rewrite systems work at a fundamental level, and by creating a rewrite system from which other rewrite systems may be constructed at a basic level. We will show that a rewrite system can represent mathematics and foundational aspects of physics and can lead to a fundamental basis for quantum computing. While this system can be encapsulated, for convenience, in a computer program written in a high level language, that program must be recognised as being different from the rewriting mechanism which it represents. It is simply a way of realising this using existing knowledge, while the originating mechanism represents 'computer language' at a much more fundamental level. The application to physics is particularly important because it is a strong test of the worth of a fundamental idea. Mathematics can be structured on fundamental principles in a large variety of ways, but physics has to survive the test of observation and experiment under many different conditions. The mathematical foundations of physics may also be expected to provide a route to understanding the principles important in the foundations of quantum computing.

In a recent paper,[1] Deutsch et al state that, 'Though the truths of logic and pure mathematics are objective and independent of any contingent facts or laws of nature, our *knowledge* of these truths depends entirely on our knowledge of the laws of physics.' According to these authors we have been forced by 'recent progress in the theory of computation', 'to abandon the classical view that computation, and hence mathematical proof, are purely logical notions independent of that of computation as a physical process'. Mathematical structures, however autonomous, 'are revealed to us only through the physical world'. We would go further and state that that mathematical structure which is most fundamental in understanding the physical world is also likely to be the structure which is most fundamental to understanding mathematics itself.

The key concepts here seem to be those of nothingness and duality. A well-known science writer, Peter Atkins, has said of physical matter that 'the seemingly something is elegantly reorganized nothing, and … the net content of the universe is … nothing'.[2] Similar comments have been made by many others, while the mathematical textbook-writer Nicholas Young has noted that 'the idea of duality pervades mathematics'.[3] According to our reasoning, however, it is not just matter and the universe that appear to be nothing, but the entire conceptual scheme of which these are merely components.[4-7] In principle, *nihil ex nihil fit*, and the way to preserve the overall nothingness is via duality. A key step in rewriting is the fact that there is an initial state. Here, we present a string representation of 0. We begin with the idea that only 0 is unique. Everything that is not 0 is undefined. In rewriting, we will start with an argument denying that we have a non-0 starting-point. We assume that we are not entitled to posit anything other than 0, and that we are force to rewrite when we start from any other position. In the process we will stress the significance of the concept of hierarchy, and of the difference between recursion and iteration.

Traditionally computer rewrite (production) systems involve objects defined in terms of symbols representing characters drawn from a finite alphabet, and a series of states. To move from state to state we apply a finite set of rules – rewrite rules or productions – to a string of the symbols that represents the current state of the complex object. Some stopping mechanism is defined to identify the end of one state and the start of the next (for example we can define that for each symbol or group of symbols in a string, and working in a specific order, we will apply every rule that applies). It is usual in such systems to halt the execution of the entire system if there is no change in the string generated or if the changes are cycling, or after a specified number of iterations. Differing stopping mechanisms determine different families of rewrite systems, and in each family, alternative rules and halting conditions may result in strings representing differing species of object. Allowing new rules to be added dynamically to the existing set and allowing rules to be invoked in a stochastic fashion are means whereby more complexity may be introduced. Examples of various types of rewrite system are given by von Koch (1905)[8], Chomsky (1956)[9], Naur et al (1960)[10], Mandelbrot (1982)[11], Wolfram (1985)[12], Prusinkiewicz and Lindenmayer (1990)[13], among others.

In this paper we show how a universal alphabet that encompasses duality and nothingness can be developed using a universal rewriting system. We examine two methods by which the elements of this alphabet may be discovered. One of these methods yields an infinite number of subset alphabets each of which has properties that can be exploited, for example using further rewrite systems based on the subset alphabet.

## 2    An infinite universal alphabet and universal rewrite systems

If we relax the rules regarding the finiteness of the characters in the alphabet(s) and the number of states, but continue to assume the rest of the constraints described above, a more universal rewrite system is defined. Such a system has alphabets as its complex objects and subset alphabets (all the symbols so far delivered) as states. For this to remain a rewrite system an initial state (that can be re-written) must exist, and for it to be universal there must be, we conjecture, a minimum of two rewrite rules (productions). One of these, *create,* delivers a new symbol at each invocation. We use the term symbol here because what is delivered may be a single character of the alphabet, a subset alphabet, or indeed an entire alphabet. The second rule, *conserve*, examines all symbols currently in existence to ensure that no anomalies exist as a consequence of bringing the new one into existence.

With such a minimum universal rewrite system, the initial state (usually called the ù-state) must contain at least one symbol that we can use to identify that the universe is empty. However, any symbol we choose is immediately (and simultaneously) a symbol, a character of the final alphabet, a subset alphabet and full alphabet in its own right. We choose, arbitrarily, the single symbol 0 (zero), and set it as the string representing the complex object in the ù-state {0}. We are obliged to make an arbitrary choice here because we cannot use *create* without the ù-state – the minimum rewrite system condition for a universal system. If we were to use *conserve* now it would simply return that 0 is unique, fixed, and consistent and no change from the ù-state would be generated. We must therefore now invoke *create* supplying the ù–state as parameter, or source, string.

If we presume that *create* is an algorithm with stopping criteria, it returns a result target string containing a new symbol. If the paradigm for the algorithm were recursive, the resulting symbol (we use E) would represent every character of the alphabet at the first step. To create any refining character, a specific $e_x$, using the recursive paradigm would be impractical because of the implied infinity and storage requirement. We may not use an iterative paradigm at this stage because we would have to supply an upper limit and/or need to identify which of the infinite characters we are creating. Both of these actions require a character not yet in the character set (alphabet) we have so far defined.

The pair of symbols, the string {0, E} is our new object (alphabet) which we now submit to *conserve* which examines every combination of symbols:

**Table 1**

|   | 0  | E  |
|---|----|----|
| **0** | 00 | 0E |
| **E** | E0 | EE |

We note that 00, the 'transition' from 0 to 0, conserves 0. The combination 0E is the transition from 0 to E and is balanced, for all E, by its conjugate partner E0 which is the transition back from E to 0, thereby conserving 0. The combination EE, the transition from every symbol E to every other, is anomalous and must be returned by *conserve* as unexplained or 'inconsistent' as it does not appear to conserve 0. However, at infinity all transitions represented by EE will have been examined, EE will be declared 'nilpotent' in that it delivers 0, and we will be left with three generic combinations:

(00, 0E, E0)

However, it is impractical to use the recursive version of *conserve* to examine further the elements of E because of the implied infinite number of iterations.

We return to the *create* process and accept that we must postulate symbols $Ä_a$, $Ä_b$, ... $Ä_n$ drawn from E such that they are in an arbitrary ordinal sequence. We note that there is an infinite number of such sequences because choice of $Ä_a$ is arbitrary. However, we may now use an iterative paradigm for *create* and because *n* is specified, an iterative (or recursive) *conserve* can be constructed. However, at the end of each invocation we are presented with a symmetrical table of transitions that represent the simplest set of properties for the current set of *n* symbols (Table 2).

**Table 2**

|   | 0 | $Ä_a$ | $Ä_b$ | $Ä_c$ | ... | $Ä_n$ |
|---|---|---|---|---|---|---|
| **0** | 00 | $0Ä_a$ | $0Ä_b$ | $0Ä_c$ | | $0Ä_n$ |
| **$Ä_a$** | $Ä_a 0$ | $Ä_a Ä_a$ | $Ä_a Ä_b$ | $Ä_a Ä_c$ | | $Ä_a Ä_n$ |
| **$Ä_b$** | $Ä_b 0$ | $Ä_b Ä_a$ | $Ä_b Ä_b$ | $Ä_b Ä_c$ | | $Ä_b Ä_n$ |
| **$Ä_c$** | $Ä_c 0$ | $Ä_c Ä_a$ | $Ä_c Ä_b$ | $Ä_c Ä_c$ | | $Ä_c Ä_n$ |
| **⋮** | | | | | | |
| **$Ä_n$** | $Ä_n 0$ | $Ä_n Ä_a$ | $Ä_n Ä_b$ | $Ä_n Ä_c$ | | $Ä_n Ä_n$ |

The $Ä_a$ row and $Ä_a$ column illustrate the conjugate pair structure observed earlier. The remaining cells of Table 2 identify explicitly each Ä symbol to Ä symbol transition observed generically in Table 1. Off diagonal there are symmetrical conjugate pairs, for example when $n = b$ there are three such cancelling pairs and six when $n = c$. The diagonal cells of the table contain transitions from each symbol to itself and do not cancel out in this way.

We now invoke the *conserve* process noting that it does not define the transition property but merely identifies those novel transition combinations that appear not to conserve 0. When $n = a$, the symbol $Ä_a$ is added to the alphabet and the transition $0Ä_a$ is introduced. We need $Ä_a0$ (and the idea that this is a conjugate form) to conserve 0. However, this leaves the combination $Ä_aÄ_a$ unexplained (novel) and to conserve 0 we must conjecture that whatever it is, is balanced by whatever is to come – or both are 'nilpotent' in the sense introduced above. To discover this we invoke *create* to add a new symbol to the alphabet which then defines (arbitrarily) the $n = b$ row and column. At $n = b$ (in *conserve*) we continue to require the conjugate explanation for all off diagonal elements in the table. In addition, we have non-0 to non-0 symbol transitions, each of which has a cancelling conjugate, and which must ultimately yield a symbol already in the alphabet. However, when these transitions are explained we still have $Ä_bÄ_b$ as novel, and require the method of explaining the novelty used earlier. We see that at every invocation of *conserve* we define the need for an additional symbol, delivered by *create* – it is inherent that both processes are obligatory. Other processes may now be conjectured within the rewrite system that impart meaning to 'transition' and also to each transition from $Ä_n$ to $Ä_n$; however, in each case all of what is to come must balance the $Ä_nÄ_n$ in the diagonal position. 'Balance' in this explanation assumes that the 00 transition yields 0, however, we could consider it to yield a conjugate of some form. Where this is the case we may consider each newly created diagonal element as 'balancing' that conjugate by delivering the unconjugated form. In each case the new symbol created carries the entire subset alphabet.

The properties and symbols emerge from the application of the two rewrite rules and would have been equally valid for any of the infinite alternative selections. Significantly, since the ultimate aim is to recover the zero state through an infinite series of processes, the emergence should be seen as being of a *supervenient* nature, that is, without temporal connotation. Furthermore, the symbol delivered at each step has all the properties of all the symbols previously delivered, and in a hierarchical and orthogonal fashion.

Finally, we note that the symbol 0, the existence of the ù-state, and the processes *create* and *conserve* are outside the rewrite system in that they must exist before the system can function. If we can allow these assumption, we may also presume the existence of some natural machine that will deliver, for a set of appropriate rewrite rules, a corresponding alphabet where the symbols themselves map to specific rules.

## 3   Mathematics

It has become a standard procedure to derive mathematical structures from the process of counting using the natural numbers, 1, 2, 3, …, and then progress by successively extending the set to incorporate negative, rational, algebraic, real, and complex numbers, before proceeding to higher algebraic structures involving, say, quaternions, vectors, Grassmann and Clifford algebras, Hilbert spaces, and even higher structures. However, to begin mathematics with the integers, though natural to our human perceptions, is to start from a position already beyond the beginning. The integers are loaded with a mass of assumptions about mathematics. They are not fundamentally simple but already contain packaged information about things beyond the integer series itself. This makes them a convenient codification of mathematics, but not a simplified starting-point. The number 1 is not the most obvious initial step from 0 because it contains, for example, the notion of discreteness, as well as ordinality. In addition, there is no obvious route of progression from natural numbers to reals. It would seem to be more logical, in terms of rewrite procedures, to begin with the real 'numbers'.

However, when we first conceive of the real 'numbers', they are not numbers at all. They are not related to anything concerned with counting, because counting does not yet exist. The set of reals ($\hat{A}$) is simply one of things unspecified. Our starting-point must be non-specific, and could be anything. We don't define it at all, not even as a set. In terms of the rewrite procedures we have adopted, such an assumption of any non-zero category must immediately lead to the return to zero, which, in mathematical terms, becomes equivalent to supposing a 'negative' category or 'conjugate' corresponding to the original assumption. (In terms of Table 2, this is the recognition that $Ä_a Ä_a$ leads to the creation of the new symbol $Ä_b$.) At this point we have created ordinality, though not yet counting, as there is no discreteness or anything fixed involved in the procedure.

It is the next application of the create procedure ($Ä_b Ä_b \rightarrow Ä_c$) which leads to the number system as we know it, for now we have an undifferentiated 'set' of possible origins for the 'negative' ordinal category or conjugate. We describe these as complex forms ($\mathbb{C}$), and each must have its own conjugate. In mathematical terms, the complex category remains completely undefined in respect to the real category, and has no ordinal relation to it. There are infinitely possible or indefinitely possible systems that are represented by the mathematical $\mathbb{C}$, even for a seemingly specified real category. It is only when we express this fact in the next creation stage that we are able to begin to extend ordinality towards enumeration, for this stage leads to what become mathematical 'combinations' of complex categories. We find here that to every conceivable $\mathbb{C}$, e.g. $\mathbb{C}$, $\mathbb{C}$, $\mathbb{C}$, …, there are indefinitely possible (commutative) combinations leading to the original real category (e.g. $\mathbb{C}\mathbb{C} \times \mathbb{C}\mathbb{C} = \hat{A}$), but very definite (anticommutative) ones leading to the conjugate (e.g. $\mathbb{C}\mathbb{C} \times \mathbb{C}\mathbb{C} = -\hat{A}$).

These alternative possibilities relate to the respective mathematical structures which we call Grassmann and Hamilton algebras. The Grassmann algebra leads to the infinite Hilbert vector spaces, while the Hamilton algebra is responsible for the cyclic system of quaternions. It is the cyclicity of the latter which introduces discreteness or closure, and the concept of 'unity'. We can choose the default position of taking the conjugate combination to create a regular ordinal sequence. We now find that only 'one' independent $\mathbb{C}$-type concept (say $\mathbb{C}$) is associated with each conceivable $\mathbb{C}$, and we can sequence the terms ordinally by choosing indistinguishability between the $\mathbb{C}$s in every conceivable respect. So the sequence, although arbitrary, becomes a series of integral binary enumerations, which we can also apply to ordinality in the real categories. With the reals, integers, and complexity as fundamental aspects of the system, the remaining mathematical number categories (and higher algebras) can be defined by applying the ordinality condition in a variety of ways, as in conventional mathematics. No new principle is required.

In effect, the hierarchical and orthogonal mathematical structure suggested by the rewrite mechanism is the following:

| | | |
|---|---|---|
| $\hat{A}$ | undefined | $\ddot{A}_a$ |
| $\hat{A}, -\hat{A}$ | conjugation | $\ddot{A}_b$ |
| $\hat{A}, -\hat{A}, ℂ, -ℂ$ | complexification | $\ddot{A}_c$ |
| $\hat{A}, -\hat{A}, ℂ, -ℂ, ℂ, -ℂ, ℂℂ, -ℂℂ$ | dimensionalization | $\ddot{A}_d$ |
| $\hat{A}, -\hat{A}, ℂ, -ℂ, ℂ, -ℂ, ℂℂ, -ℂℂ,$ | repetition | $\ddot{A}_e$ |
| $ℂ, -ℂ, ℂℂ, -ℂℂ, ℂ ℂ, -ℂ ℂ,$ | | |
| $ℂℂ ℂ, -ℂℂ ℂ$ | | |

The subset alphabets at each step represent all those, including $\hat{A}, -\hat{A}$ which are generated by operating on themselves:

$(\hat{A}) \times (\hat{A}) = (\hat{A})$
$(\hat{A}, -\hat{A}) \times (\hat{A}, -\hat{A}) = (\hat{A}, -\hat{A})$
$(\hat{A}, -\hat{A}, ℂ, -ℂ) \times (\hat{A}, -\hat{A}, ℂ, -ℂ) = (\hat{A}, -\hat{A}, ℂ, -ℂ)$
$(\hat{A}, -\hat{A}, ℂ, -ℂ, ℂ, -ℂ, ℂℂ, -ℂℂ) \times (\hat{A}, -\hat{A}, ℂ, -ℂ, ℂ, -ℂ, ℂℂ, -ℂℂ)$
$\qquad = (\hat{A}, -\hat{A}, ℂ, -ℂ, ℂ, -ℂ, ℂℂ, -ℂℂ)$, etc.

From this structure, and from the general rule that a character set operating on itself or any set or symbol contained within it produces itself, we may obtain rules between the individual characters, $\hat{A}, ℂ$, etc., of the form:

$\hat{A} \times \hat{A} = -\hat{A} \times -\hat{A} = \hat{A}$
$\hat{A} \times -\hat{A} = -\hat{A} \times \hat{A} = -\hat{A}$
$\hat{A} \times ℂ = ℂ \times \hat{A} = ℂ$
$ℂ \times ℂ = -ℂ \times -ℂ = -\hat{A}$
$ℂ \times -ℂ = -ℂ \times ℂ = \hat{A}$
$ℂ \times ℂ = -ℂ \times -ℂ = -\hat{A}$
$ℂℂ \times ℂℂ = -ℂℂ \times -ℂℂ = -\hat{A}$     closed (anticommutative)
$ℂℂ \times ℂℂ = -ℂℂ \times -ℂℂ = \hat{A}$     unlimited (commutative)

The choice between the last two procedures is not determined by the algebra. Both are true infinitely and an infinite number of each would be contained within E. However, since we consider the generating mechanism to be supervenient, we can structure it to default at the first option, and so generate an infinite number of identically closed systems, from which we derive an infinite integral sequence.

Here we establish for the first time the meaning of both the number 1 and the binary symbol 1 as it appears in classical Boolean logic. We identify the logical 1 as potentially a conjugation state of 0, that is, a subset alphabet defined within the system. Alternative systems of units will be possible, where they can related by a mapping to the overall structure, for example the negative unit system developed by Santilli, with its powerful applications.[14]

We can, also, for our convenience, use the integral ordinal sequence established with dimensionalization to restructure the subset alphabets as a series of finite groups, the order of which doubles at every stage, producing an ordinal binary enumeration. The succession, allowing for conjugation (±) within each group, becomes:

| | |
|---|---|
| order 2 | real scalar |
| order 4 | complex scalar (pseudoscalar) |
| order 8 | quaternions |
| order 16 | complex quaternions or multivariate vectors |
| order 32 | double quaternions |
| order 64 | complex double quaternions or multivariate vector quaternions |

Defining closure in terms of enumeration further allows us to understand $\hat{A}$ in terms of the set of real numbers (defined by the Cantor continuum), with + and × now understood as the processes of mathematical addition and multiplication. The dimensional or constructible 'real' numbers represented by terms such as $\mathbb{C}\mathbb{C}$ (with countable units squaring to 1) would then be equivalent to those of Robinson's non-standard analysis or Skolem's non-standard arithmetic. From this particular interpretation, it is possible to develop new types of mathematics by combining different aspects of the overall structure in novel ways, as has been the usual procedure in mathematics, and we conjecture that all branches of mathematics that can conceivably exist may be generated by procedures internal to this structure.

There are, effectively, only three processes at work: conjugation, which produces the alternative + and – values; complexification, which introduces a new complex factor of the form $\mathbb{C} = i$; and dimensionalization, which introduces a complementary complex factor of the form $\mathbb{C} = j$, converting the $i$ into an element of a quaternion set. The sequence proceeds through an infinite series of quaternionic structures by repeated processes of complexification and dimensionalization. (It is significant that further applications of conjugation would not affect the structure of the elements in the groups.)

In terms of 'units' (once we have established their existence), we could express the structures in the form:

order 2   $\pm 1$
order 4   $\pm 1, \pm i_1$
order 8   $\pm 1, \pm i_1, \pm j_1, \pm i_1 j_1$
order 16  $\pm 1, \pm i_1, \pm j_1, \pm i_1 j_1, \pm i_2, \pm i_2 i_1, \pm i_2 j_1, \pm i_2 i_1 j_1$
order 32  $\pm 1, \pm i_1, \pm j_1, \pm i_1 j_1, \pm i_2, \pm i_2 i_1, \pm i_2 j_1, \pm i_2 i_1 j_1,$
          $\pm j_2, \pm j_2 i_1, \pm j_2 j_1, \pm j_2 i_1 j_1, \pm j_2 i_2, \pm j_2 i_2 i_1, \pm j_2 i_2 j_1, \pm j_2 i_2 i_1 j_1$
order 64  $\pm 1, \pm i_1, \pm j_1, \pm i_1 j_1, \pm i_2 i_1, \pm i_2 i_1, \pm i_2 j_1, \pm i_2 i_1 j_1,$
          $\pm j_2, \pm j_2 i_1, \pm j_2 j_1, \pm j_2 i_1 j_1, \pm j_2 i_2, \pm j_2 i_2 i_1, \pm j_2 i_2 j_1, \pm j_2 i_2 i_1 j_1$
          $\pm i_3, \pm i_3 i_1, \pm i_3 j_1, \pm i_3 i_1 j_1, \pm i_3 i_2, \pm i_3 i_2 i_1, \pm i_3 i_2 j_1, \pm i_3 i_2 i_1 j_1,$
          $\pm i_3 j_2, \pm i_3 j_2 i_1, \pm i_3 j_2 j_1, \pm i_3 j_2 i_1 j_1, \pm i_3 j_2 i_2, \pm i_3 j_2 i_2 i_1, \pm i_3 j_2 i_2 j_1, \pm i_3 j_2 i_2 i_1 j_1$

Usually, of course, $i_1 j_1$ would be written $k_1$, but no new independent unit is created by this notation. An alternative expression could be in terms of multiplying factors:

order 2   $(1, -1)$
order 4   $(1, -1) \times (1, i_1)$
order 8   $(1, -1) \times (1, i_1) \times (1, j_1)$
order 16  $(1, -1) \times (1, i_1) \times (1, j_1) \times (1, i_2)$
order 32  $(1, -1) \times (1, i_1) \times (1, j_1) \times (1, i_2) \times (1, j_2)$
order 64  $(1, -1) \times (1, i_1) \times (1, j_1) \times (1, i_2) \times (1, j_2) \times (1, i_3)$ ,

with the series repeating for an endless succession of indistinguishable $i_n$ and $j_n$ values. It is the potentially infinite sequence of $i_n$ values, with commutativity between $i_m$ and $i_n$ or $j_n$ ($m \neq n$), which creates the possibility of a Grassmann or infinite-dimensional vector algebra, while the anticommutativity between $i_n$ and $j_n$ ensures the finite- and, specifically, three-dimensionality of each of the quaternion systems. The commutativity of $i_m$ and $i_n$ is equivalent to defining $(i_m i_n)^2$ as 1, while the anticommutativity of $i_n$ and $j_n$ defines $(i_n j_n)^2$ as the conjugate, or –1. It is notable that there is no such thing, in principle, as a pure complex number, only an incomplete representation of a quaternion set.

The order 16 group is of special interest as creating what is effectively a 'real' dimensional structure of the kind observed in normal 3-dimensional vector space. The components, $\pm 1, \pm i_1, \pm j_1, \pm i_1 j_1, \pm i_2, \pm i_2 i_1, \pm i_2 j_1, \pm i_2 i_1 j_1$, could be more conveniently rearranged and written in the form $\pm 1, \pm i, \pm \mathbf{i}, \pm \mathbf{j}, \pm \mathbf{k}, \pm i\mathbf{i}, \pm i\mathbf{j}, \pm i\mathbf{k}$, where $\pm 1, \pm i$, become the respective scalar and pseudoscalar, and $\mathbf{i}, \mathbf{j}, \mathbf{k}$, and $i\mathbf{i}, i\mathbf{j}, i\mathbf{k}$ the respective vector and pseudovector terms of the multivariate algebra, explored by Hestenes and others,[15,16] and applied by them to the algebra of physical space and time, to generate electron spin as a natural consequence of spatial three-dimensionality. This is the algebra of Pauli matrices, in which the 'total' product of two multivariate vectors **a** and **b** is of the form $\mathbf{a}.\mathbf{b} + i\, \mathbf{a} \times \mathbf{b}$, and the 'total' products of the vector units is of the form $\mathbf{ii} = \mathbf{jj} = \mathbf{kk} = 1$; and $\mathbf{ij} = -\mathbf{ji} = i\mathbf{k}$; $\mathbf{jk} = -\mathbf{kj} = i\mathbf{i}$; and $\mathbf{ki} = -\mathbf{ik} = i\mathbf{j}$.

The order 16 group also (if we are to retain the maximum indistinguishability by avoiding octonion-type nonassociativity) is the point at which the extension of the sequence becomes one of repetition, and so a complete specification of an interative procedure could be made by using the groups of order 2, 4, 8 and 16. Taken as independent entities, these may be combined in the group of order 64, using the symbols $\pm 1$, $\pm i$, $\pm \mathbf{\mathit{i}}$, $\pm \mathbf{\mathit{j}}$, $\pm \mathbf{\mathit{k}}$, $\pm \mathbf{i}$, $\pm \mathbf{j}$, $\pm \mathbf{k}$, to represent the respective units required by the scalar, pseudoscalar, quaternion and multivariate vector groups. This takes on physical significance when we realize that the algebra of this group is that of the gamma matrices used in the Dirac equation – the quantum equation determining the behaviour of the most fundamental components of matter – and that these matrices may be represented as the terms $\mathbf{\mathit{k}}$, $i\mathbf{i}$, $i\mathbf{j}$, $i\mathbf{k}$, $i\mathbf{\mathit{j}}$, whose binomial combinations are sufficient to generate the entire group.[17,18] So the minimal mathematical structure which most closely corresponds to the 'unit' required to generate the iterative procedure of our rewrite mechanism would appear to be the one which is most significant to physics at the foundational level. Mathematical analysis also shows that the reduction of the group elements to a smaller number of composite generating units is only possible in a pentad or 5-fold form either identical or isomorphic to the Dirac matrices. It is significant for physics that this creates a naturally broken symmetry.

**4   Physics**

Each of the processes involved in the generation of the sequence of mathematical structures by the rewrite mechanism – conjugation, complexification, and dimensionalization – would appear to have a realization in physics, which seemingly contrives to use the minimum possible structure for returning to zero without privileging any of the component processes. A structure previously proposed as foundational to physics suggests that the only truly fundamental parameters are space, time, mass(-energy) and charge, which are respectively represented as multivariate vector, pseudoscalar, real scalar and quaternion.[4-7] The quaternion nature of charge is indicated by its existence in three types (electric, strong and weak), and the fact that interactions between identical charges are of opposite sign to those between identical masses. The parameters also have an internal group symmetry, which, for the purposes of this discussion, can be expressed in the following form:

| | | | |
|---|---|---|---|
| **space** | nonconjugated | real | dimensional |
| **time** | nonconjugated | complex | nondimensional |
| **mass** | conjugated | real | nondimensional |
| **charge** | conjugated | complex | dimensional |

Conjugated here is equivalent to conserved, so a positive charge (or source of mass-energy) cannot be created without also creating a negative one. Significantly, only the (3-)dimensional quantities, space and charge, are countable, and, physically, one cannot imagine a mechanism for dividing the units in a single dimension. (This is why time is physically irreversible and mass-energy is physically unipolar; neither quantity allows a discontinuity or zero state.) In addition, the mathematical processes which allow for the continual recreation of new non-integral structures in 1-to-1 correspondence with the integers would be inconceivable in a system without dimensionality. As in conventional mathematics, two versions of the 'real' numbers are required: the uncountable ones of the Cantor continuum and standard analysis (for mass), and the countable ones of the Löwenheim-Skolem arithmetic and Robinson's non-standard analysis (for space).

If the combination of these parameters, or of the real scalar, pseudoscalar, multivariate vector, and quaternion units by which they are realized, is to become itself a 'unit' of the rewrite procedure, we should expect to find some degree of 'closure' or cyclicity, parallel to that which produces the pure quaternion system. Now, a fundamental aspect of the quaternion algebra, which, in our system, introduces discreteness, enumeration, or countability, is that it is anticommutative, and it is this very anticommutativity which causes the cyclicity which leads to discreteness. It is significant, in this context, that the presence of anticommutativity allows physics to create a more direct route to the zeroing or conjugation of an act of 'creation', at the level of the 64-element Dirac algebra. In parameterizing the physical world using this algebra, we create a structure which zeros itself by being a nilpotent or square root of zero, so producing a cyclicity at a higher level which incorporates the whole range of procedures required for the rewrite mechanism. The next stage is then simply to make infinitely or indefinitely many applications of this closed system or 'unit' structure to construct the entire physical universe, in the same way as we iterate applications of the quaternion system to construct a system of mathematics.

The physical system which provides this 'unit' structure is the fermion wavefunction, in the most general (quantum field) version of the Dirac equation, and its algebraic realization requires the gamma matrices, or the order 64 multivariate vector-quaternion group of our series. A free fermion wavefunction, obeying the Pauli exclusion principle, can be written in the form

$$y = (\pm kE \pm ii\ \mathbf{p} + ij\ m)\ e^{-i(Et - \mathbf{p}\cdot\mathbf{r})}\ ,$$

where $\mathbf{p}$ incorporates a multivariate vector unit.[3,4] The Dirac equation,

$$\left(\pm ik\frac{\partial}{\partial t} \pm i\nabla + ijm\right)(\pm kE \pm ii\ \mathbf{p} + ij\ m)\ e^{-i(Et - \mathbf{p}\cdot\mathbf{r})} = 0\ ,$$

now specifies the relationship between the nonconserved or nonconjugated quantities, represented by the differential operator acting on the exponential term, and the conserved or conjugated quantities, represented by the nilpotent operator ($\pm kE \pm ii\ \mathbf{p} + ij\ m$). In principle, this reduces to

$$(\pm kE \pm ii\ \mathbf{p} + ij\ m)\ (\pm kE \pm ii\ \mathbf{p} + ij\ m) = E^2 - p^2 - m^2 = 0\ ,$$

which applies even when the fermions are no longer 'free'. The conservation laws incorporated into the nilpotent operator ($\pm kE \pm i\textbf{i}\,\textbf{p} + ij\,m$) include those of mass-energy and the three types of charge, information on the latter being carried by the orientation, direction and magnitude of the angular momentum. It is these conservation laws, defined against the nonconservation or variation of space and time, which determine the behaviour of physical systems.

At the quantum level, the physical universe appears to be composed entirely of fermionic or antifermionic wavefunctions of this kind or of combinations of them. Antifermionic wavefunctions merely reverse or conjugate the sign of $kE$ in fermion wavefunctions, producing equally nilpotent terms such as ($\mp kE \pm i\textbf{i}\,\textbf{p} + ij\,m$), while bosonic wavefunctions are nothing other than combinations of the two. The 'universe', as described by physicists, is essentially an entanglement of all possible nilpotent states. Significantly, no nilpotent can be identical to any other; each must be unique, with $E$, $p$ and $m$ being unspecified real numbers. (The Berry phase, as manifested in the quantum Hall, Aharonov-Bohm or Jahn-Teller effects, is effectively a realisation of the equivalent antifermionic wavefunction in a fermion's physical 'environment', and is, in this sense, a unique signature.[19]) Ordinality is preserved, but enumeration is reserved for the nilpotent units rather than their component parts. However, the creation of the nilpotent operator is equivalent to the process of quantization of $E$, $p$ and $m$, which thus become 'dimensionalized', while, the quaternions specifying the weak, strong and electric charges ($k$, $i$, $j$) become distinguished by being attached, respectively, to scalar, multivariate vector and pseudoscalar operators. With this built-in degree of symmetry-breaking, we are on the way to understanding important aspects of fundamental physics.

## 5  Uniqueness, qubits and quantum computing

The nilpotent algebra used in the Dirac formalism, together with its infinite Hilbert space expansion, provides a mathematics of uniqueness previously unexplored. Mathematics is normally structured on the notion that its units are capable of repeated application, but the infinite nilpotent algebra is structured on the idea that its units, though variable, cannot be repeated. This is because a superposition of any two identical nilpotents, such as the Hilbert space formalism requires, will be automatically zero. The only way to make a nonzero universe out of these is for each to be unique. This is manifested physically as Pauli exclusion. By using a series of operators that can be repeated, conventional mathematics loses some of the information which is potentially available; an algebraic structure based on unique operators may be expected to produce an entirely new set of mathematical results. Significantly, the uniqueness, though the result of an infinite superposition, is manifested within any finite set of nilpotent operators. The nilpotent algebra thus allows us to use the iterative procedure to represent a recursive system.

The formalism is only possible because the terms $E$, $\textbf{p}$ and $m$, like the original parameters time, space and mass, from which they were derived, have the full range of real number values. In principle, then, each individual nilpotent can be unique; and must be if, as we believe, the entire universe can be structured as a superposition of fermionic states, with any nonuniqueness in the components producing immediate zeroing. The generating algebra which we have created by our rewrite mechanism can then be extended to infinity, through the physical property of fermionic wavefunctions being nonlocally connected throughout the entire universe. In principle, it is the mathematical interconnectedness of the nilpotent operators that allows us to group its components as a 'unit' of an even higher algebra, which may be in the form of the conventional complex Hilbert space or the equivalent geometric algebra as demonstrated by Matzke (or even a complex version of the latter).[20]

These algebras provide the doubling mechanism provided in our foundational algebraic structure by terms of the form $(1, i_n)$. In effect, the fermionic nilpotents become isomorphic to the fundamental unit of quantum information, or qubit, composed of two orthogonal vectors and their superposition states. So, taking the tensor product of every qubit expands the space exponentially, exactly as in our mathematics of duality. In principle, also, such fermionic qubits would be uniquely labelled, as required for quantum computing, and, theoretically, the fermionic states could be identified by a manifestation of the Berry phase, such as the quantum Hall effect. (Alternatively, an ideal Bose-Einstein condensate would consist of fermionic nilpotents differing only by their opposite spin states, or sign of **p** in the nilpotent formalism.) Deutsch, in his classic foundational paper on quantum computing,[21] states that any physical process can be modelled perfectly by a quantum computer. This, according to our understanding, is because a quantum computer is ultimately described in terms of the same units as real physical processes. The physical universe is the set of all possible quantum computers.

## 6  Conclusion

The aim of this paper is not to deny the fundamental nature of mathematics or physics, but to capture the way they operate at a more fundamental level, and, by doing so, to gain an extra power of understanding and manipulation. Mathematics can be shown to be constructible using this mechanism, with an order which is more coherent than one produced by starting with integers. By rejecting the 'loaded information' that the integers represent, and basing our mathematics on an immediate zero totality, we believe that we are able to produce a mathematical structure which has the potential of avoiding the incompleteness indicated by Gödel's theorem. (Conventional approaches, based on the primacy of the number system, have necessarily led to the discovery that a more primitive structure cannot be recovered than the one initially assumed.) From this mathematical structure, we have been able to develop an insight into how physics works, and using this to suggest a process that leads naturally to a formulation for quantum computation.

The structure may be found relevant also to other aspects of theoretical computation especially abstract machine specification where notation and the needs of rewriting (substitution) languages are explicitly required.[22] The universal rewrite system that is proposed may be mapped to a Turing machine, very close to Turing's original assumptions where every operation 'consists of some change in the physical system consisting of the computer and his tape'.[23] and every subset alphabet can be used in such an environment. For example at a simple level the subset alphabet with conjugation alone, when appropriately wrapped, provides an exact mapping to a Boolean encoding and, when a symbolism for the conjugate character is added, maps to a ternary encoding.

A physical universe composed of a potentially infinite series of unique (but changeable) nilpotents, originating in the supervenient dualistic processes needed to maintain the zero total state, has itself all the characteristics of a Turing machine. The description of physical systems in these terms allows a mapping of Turing systems to other physical processes and suggests a novel approach to investigating such systems. Here the algebraic and rewrite structure that underlies the mapping can be used to simulate and demonstrate such systems.

In addition, infinitely parallel and serial systems are posited by the method when generating Grassmann and Hamilton type algebras. Though our own system is parallel in the first instance, and ideal for quantum computation, we have options in what we can select out from our mathematical structure, and could also have chosen a serial representation. Indeed to propose the universal alphabet in a representation that encompasses natural physics we are required to follow both a serial and iterative procedure. The structure we present has all the properties required of a universal rewrite system that can generate its own alphabet.

In addition to its immediate relevance to quantum computation and theoretical computation, to mathematics, and to physics, we believe that this approach has possible practical application in parallel computation. This is especially the case when cast as parallel agents having autonomous actions mediated by message passing within a well defined spatial and temporal set of constraints. The required properties of this processing environment are captured by the concept of a subset alphabet, and process steps and communication mechanisms are represented as rewrite rules. It is likely that this sort of parallel processing environment will have immediate application to our understanding of the complexity of biological and biotechnological systems.